\documentclass[aps,preprint]{revtex4}
\usepackage{amssymb}
\usepackage{latexsym}
\usepackage{amsfonts}
\usepackage{graphicx}
\usepackage{epsfig}
\usepackage{amsmath}
\usepackage{float}  
\usepackage{verbatim}

\begin{document}

\title{Multifactor Analysis of Multiscaling in Volatility Return
Intervals}

\author{Fengzhong Wang$^1$, Kazuko Yamasaki$^{1,2}$, Shlomo
Havlin$^{1,3}$ and H. Eugene Stanley$^1$}

\affiliation{$^1$Center for Polymer Studies and Department of Physics,
Boston University, Boston, MA 02215 USA\\$^2$Department of
Environmental Sciences, Tokyo University of Information Sciences,
Chiba 265-8501,Japan\\$^3$Minerva Center and Department of Physics,
Bar-Ilan University, Ramat-Gan 52900, Israel}

\date{22 August 2008 ~~~ wyhs.tex}

\begin{abstract}

We study the volatility time series of 1137 most traded stocks in the
US stock markets for the two-year period 2001-02 and analyze their
return intervals $\tau$, which are time intervals between volatilities
above a given threshold $q$. We explore the probability density
function of $\tau$, $P_q(\tau)$, assuming a stretched exponential
function, $P_q(\tau) \sim e^{-\tau^\gamma}$. We find that the exponent
$\gamma$ depends on the threshold in the range between $q=1$ and $6$
standard deviations of the volatility. This finding supports the
multiscaling nature of the return interval distribution. To better
understand the multiscaling origin, we study how $\gamma$ depends on
four essential factors, capitalization, risk, number of trades and
return. We show that $\gamma$ depends on the capitalization, risk and
return but almost does not depend on the number of trades. This
suggests that $\gamma$ relates to the portfolio selection but not on
the market activity. To further characterize the multiscaling of
individual stocks, we fit the moments of $\tau$, $\mu_m \equiv
\langle(\tau/\langle\tau\rangle)^m\rangle^{1/m}$, in the range of $10
< \langle\tau\rangle \le 100$ by a power-law, $\mu_m \sim
\langle\tau\rangle^\delta$. The exponent $\delta$ is found also to
depend on the capitalization, risk and return but not on the number of
trades, and its tendency is opposite to that of $\gamma$. Moreover, we
show that $\delta$ decreases with $\gamma$ approximately by a linear
relation. The return intervals demonstrate the temporal structure of
volatilities and our findings suggest that their multiscaling features
may be helpful for portfolio optimization.

\end{abstract}


\maketitle

The study of volatility has long been one of the main topics of
economics and econophysics research \cite{Mandelbrot63,Mantegna95,
Kondor99,Mantegna00,Takayasu97,Bouchaud00,Johnson03,Liu99,Plerou01}. It
is important for revealing the mechanism of price dynamics as well as
for developing strategies of investment. For example, it helps the
investor to estimate the risk and optimize the portfolio
\cite{Bouchaud00,Johnson03}. As a stylized fact of econophysics, the
volatility time series has long-term power-law correlations
\cite{Liu99,Plerou01,Ding83,Wood85,Harris86}. The temporal structure
in volatilities is complex and still regarded as an open
problem. Return interval $\tau$, also called recurrence time or
interspike interval, which is the time interval between two
consecutive volatilities above a certain threshold $q$, provides a new
approach to analyze long-term correlated time series
\cite{Altmann05,Yamasaki05,Wang06,Weber07,Vodenska08,Jung08,
Bunde04,Livina05,Lennartz08,Eichner07,Bogachev07,Wang08}. Recent
studies on financial markets
\cite{Yamasaki05,Wang06,Weber07,Vodenska08,Jung08} show that, for both
daily and intraday data, i) the distribution of scaled interval
$\tau/\langle\tau\rangle$ can be approximated by a single scaling
function, where $\langle\tau\rangle$ is the average of $\tau$. The
scaling function can also be approximated by a stretched exponential
(SE) function. ii) The sequences of the return intervals have
long-term memory which is related to the long-term correlations in the
original volatility sequences. Similar findings are observed for other
long-term correlated time series, such as climate and earthquake
\cite{Bunde04,Livina05,Lennartz08}. Also there are some related
studies on financial markets, such as first passage time
\cite{Masoliver05} and level crossing \cite{Jafari06}.

As a typical complex system, financial market is composed of many
interconnected participants and its time series is usually not of
uniscaling nature \cite{Matteo07}. Market activity such as the
intertrade time shows multiscaling in its distribution
\cite{Ivanov04,Eisler06}. Recently we suggested that the return
intervals distribution has multiscaling characteristics based on
cumulative distributions and moments of scaled intervals for 500
constituents of the Standard \& Poor's 500 index \cite{Wang08}. The
following questions are, can we detect multiscaling for a broader
market? More important, what is the reason for multiscaling in the
return intervals? Is it related to the market activity? Or is it
connected to the portfolio selection criteria such as company size,
stock risk or return?  The study of those possible relations may shed
light on the underlined mechanism of the volatility and may help
investors to optimize their portfolio.

In this paper we analyze the volatility return intervals of the entire
US stock markets. The database analyzed is the Trades And Quotes (TAQ)
from New York Stock Exchange (NYSE). The period studied is from Jan 1,
2001 to Dec 31, 2002, totally 500 trading days. TAQ records every
trade (``tick'') for all securities in the US stock markets. The stock
activity varies in a wide range, between $5$ and $65,000$ trades per
day. For constructing a minute resolution data, one need enough
records in every day and thus we choose only stocks that have at least
500 daily trades. With this criterion, we obtain 1137 stocks which are
the most traded in the market. From tick prices we set the closest one
to a minute mark as the price at that minute. The volatility is
defined the same as in Ref \cite{Wang06}. First, we compute the
absolute value of the logarithmic change of the minute price, then
remove the intraday U-shape pattern, and finally normalize the series
with its standard deviation. Therefore the volatility is in units of
standard deviations. Since the sampling time is 1 minute, a trading
day has 390 points (after removing the market closing hours), and each
stock has about 195,000 records.

The analysis with respect to several essential factors is widely used
in economics studies. For instance, company size, market return and
book-to-market value are used to model asset pricing
\cite{Fama96}. Volatilities and therefore return intervals may be
affected by many factors. Here we study how the return intervals
distribution depends on a few essential measures which characterize
different features of the stocks. The first one is the size of
company, which is a popular criterion for portfolio selection. Stocks
of different scales are preferred by investors of different types. The
size also limits the group of investors and market depth for a
stock. On the other hand, the internal organization of a company might
dramatically varies with its size. Thus, the volatility and its return
interval may be strongly influenced by this factor. The size is
usually characterized by the market capitalization, product of the
stock price and outstanding shares. Without loss of generality, we
choose the price and outstanding shares on Dec 31, 2002 to calculate
the capitalization. For the 1137 stocks, the range of capitalization
is between $2 \times 10^7$ and $2 \times 10^{11}$ dollars.

The reward and risk are basic concerns for any investment and we
therefore choose them as the next two factors. The reward is usually
measured as the average return of price while risk is measured as the
standard deviation of the return \cite{Markowitz52}. This traditional
definition of the risk is based on the Gaussian distribution of the
time series, which is not always adapted to the financial data
\cite{Bouchaud00}. Nevertheless, it characterizes the magnitude of
fluctuations and therefore the risk. To avoid the intraday pattern
\cite{Liu99,Wang06}, we calculate the return on a daily basis. The
return is the logarithmic daily price change averaged over the
two-year period (2001-2002), which varies from $-0.008$ to $0.004$ for
the 1137 stocks. The risk, standard deviation of daily returns in the
two years, ranges from $0.012$ to $0.12$. The fourth factor we study
is an activity measure, the number of trades per day. Note that the
four factors reflect different aspects of a stock. The size is for the
scale of company. The return and risk are historical price movement
tendency and variation, which are helpful for the prediction of future
price change. While the number of trades shows the activity, i.e., how
frequent a stock is traded.

\begin{figure*}
\begin{center}
   \includegraphics[width=0.65\textwidth, angle = -90]{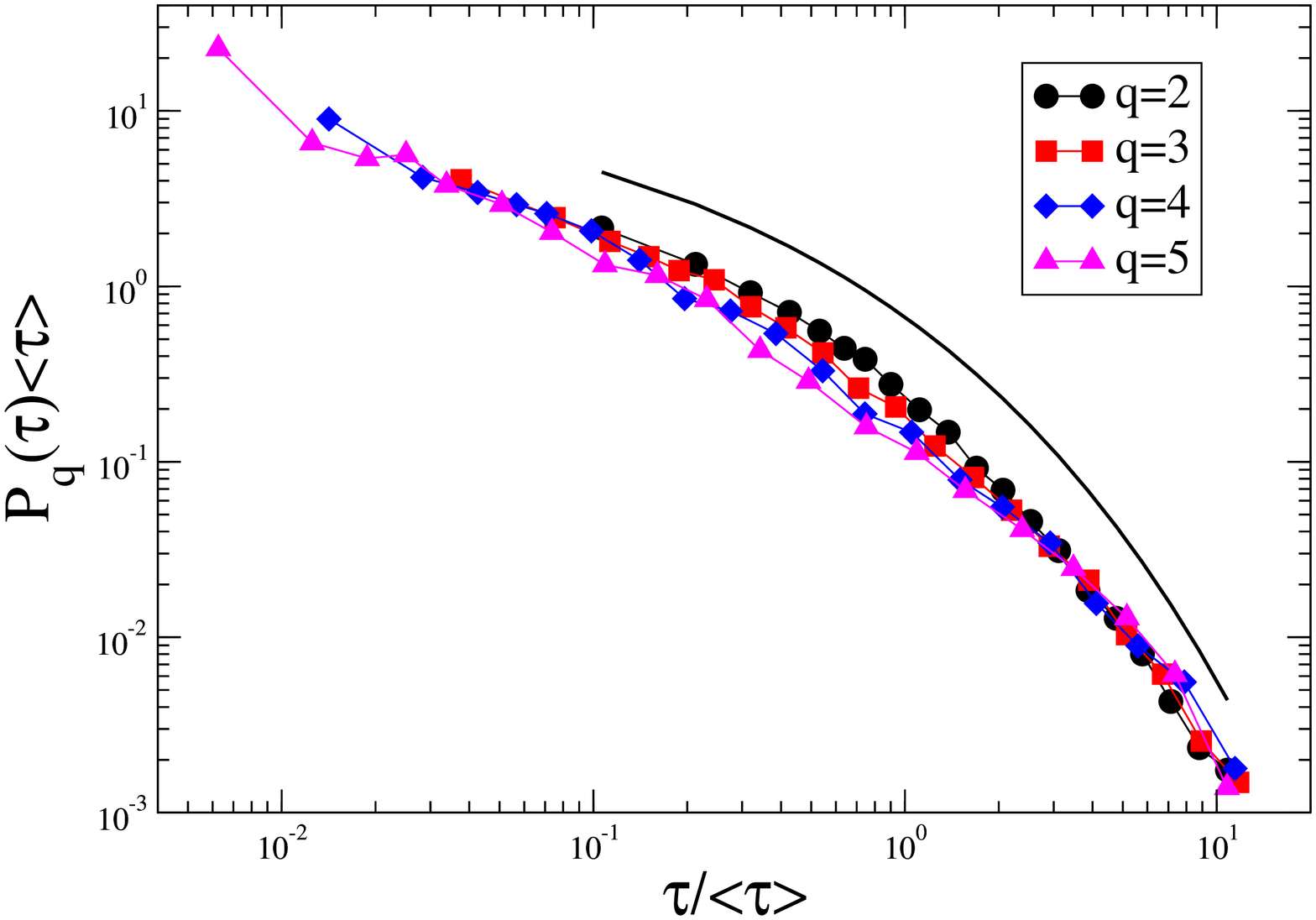}
\end{center}
\caption{(Color online) Return interval PDFs of four thresholds,
$q=2$, $3$, $4$ and $5$ for the GE stock. These four curves
approximately collapse onto a single one, and the scaling function is
approximate stretched exponential, as guided by the black curve which
is the SE fitting to the data for $q=2$ (shifted vertically for better
visibility).}
\label{Fig1}
\end{figure*}

For a volatility time series, we choose a positive value as the
threshold $q$ and find those volatilities above $q$, which are called
``events''. Note that $q$ is in units of standard deviations. Then we
calculate the time intervals $\tau$ between two consecutive events and
compose a new time series. For each threshold $q$ we have a
corresponding time series of return intervals. For financial markets,
the PDF of $\tau$, $P_q(\tau)$, is well-approximated by the scaling
function \cite{Yamasaki05,Wang06},
\begin{equation}
P_q(\tau)=\frac{1}{\langle\tau\rangle}f(\tau/\langle\tau\rangle),
\label{pdf.eq}
\end{equation}
where $\langle\cdot\rangle$ stands for the average over the data
set. The scaling function $f(x)$, where $x$ corresponds to the scaled
interval $\tau/\langle\tau\rangle$, can be approximated by a SE
function,
\begin{equation}
f(x)=c e^{-(ax)^\gamma},
\label{scaling.eq}
\end{equation} 
in consistent with other long-term correlated records
\cite{Bunde04,Altmann05}. The normalization constant $c$ and the
scaled parameter $a$ depend on the exponent $\gamma$, and thus $f(x)$
has only one free parameter \cite{Altmann05,Wang08}. When the record
has no long-term correlations, the return intervals follow as expected
an exponential distribution, i.e. $\gamma=1$. As an example, we plot
in Fig. \ref{Fig1} the PDFs of return intervals for a typical stock,
General Electric (GE). The PDFs for four values of $q$ ($q=2$ to $5$)
almost collapse onto a single curve. We also plot a SE
(Eq. (\ref{scaling.eq}) fitting of the curve for $q=2$. For small
values of $\tau$, there are some deviations from the SE
function. Eichner et al. suggested that the scaling function is
characterized by a power-law function for short time scales and a SE
function for long time scales \cite{Eichner07}. To avoid these
deviations, we analyze the scaling function only for large scales
($\tau/\langle\tau\rangle \ge 0.1$).

\begin{figure*}
\begin{center}
   \includegraphics[width=0.65\textwidth, angle = -90]{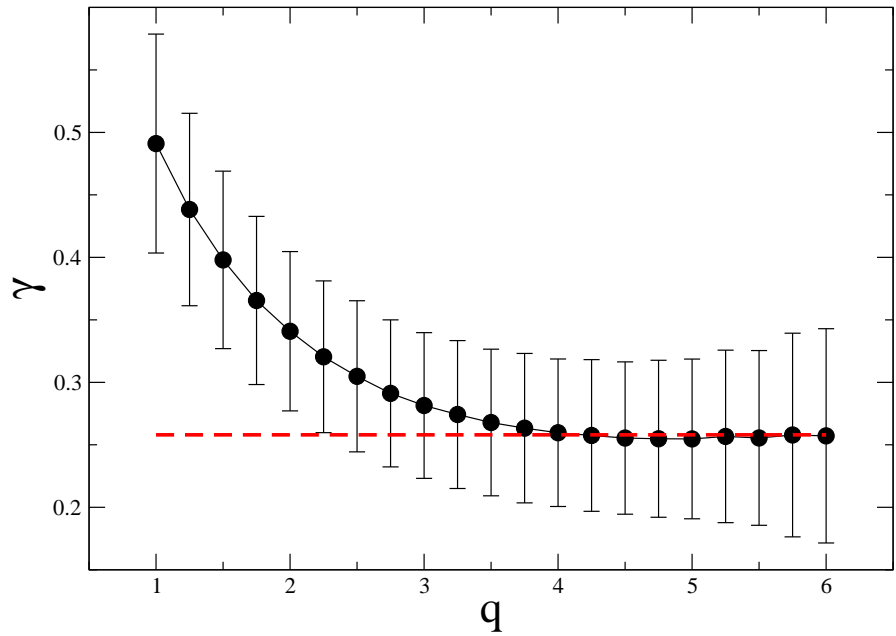}
\end{center}
\caption{(Color online) SE exponent $\gamma$ vs. threshold $q$. The
filled circles are the values of $\gamma$ averaged over 1137 most
traded stocks and the error bars are the corresponding standard
deviations. The dashed line is a guide line of $\gamma=0.26$.}
\label{Fig2}
\end{figure*}

In a recent paper \cite{Wang08}, indications of deviations from the
scaling function Eq. (\ref{scaling.eq}) were observed for the return
intervals. The cumulative distributions for different thresholds $q$
were found to systematically deviate from a single scaling function
\cite{Wang08}. This indicates that the exponent $\gamma$ may change
with the threshold $q$. To test this assumption quantitatively and
over the entire market, we compute $\gamma$ for all 1137 stocks and
plot in Fig \ref{Fig2} their averages and standard deviations (as
error bars) as a function of $q$. The values of $\gamma$ are obtained
from the least-squares fit of the scaling function,
Eq. (\ref{scaling.eq}), to the data for the range
$\tau/\langle\tau\rangle\geq0.1$ (see Fig. \ref{Fig1}). The range of
$q$ studied is from $1$ to $6$ with steps of $0.25$. We consider a
point as on outlier if its RMS error is larger than $10\%$. Totally
$730$ out of $22740$ points or $3.8\%$ of all points are
removed. Fig. \ref{Fig2} shows that the mean $\gamma$ decreases with
$q$, from $0.49$ for $q=1$ to $0.28$ for $q=3$. For large thresholds
(between $q=3$ and $6$) $\gamma$ tends to be constant (around $0.26$),
where the distribution can be regarded as close to be of uniscaling
nature. The difference in $\gamma$ between small and large thresholds
suggests multiscaling in the distribution for the whole range. The
volatility time series has long-term correlations, which can be
characterized by the exponent $\alpha$ obtained from Detrended
Fluctuation Analysis (DFA) method
\cite{Peng94,Liu99,Yamasaki05,Wang06}. Assuming the validity of the
relation between $\gamma$ in Eq. (\ref{scaling.eq}) and the long-term
correlations in the volatilities \cite{Eichner07}, $\alpha = 1 -
\gamma/2$, it follows that small volatilities have large $\gamma$ and
weak correlations, while large volatilities have small $\gamma$ and
strong correlations. Large volatilities correspond to long time scales
and small volatilities correspond to short time scales. The changes in
the value of $\gamma$ seen in Fig. \ref{Fig2} might be due to the
changes in the $\alpha$ found between short and long time scales in
the volatility records \cite{Liu99,Wang06}. The error bars shown in
Fig. \ref{Fig2} are limited for all thresholds, which indicates the
tendency is consistent for the entire market. Note that error bars for
several largest thresholds such as $q=6$ and $5.75$ are slightly
larger, probably due to the bad statistics of fewer events.

\begin{figure*}
\begin{center}
   \includegraphics[width=0.65\textwidth, angle = -90]{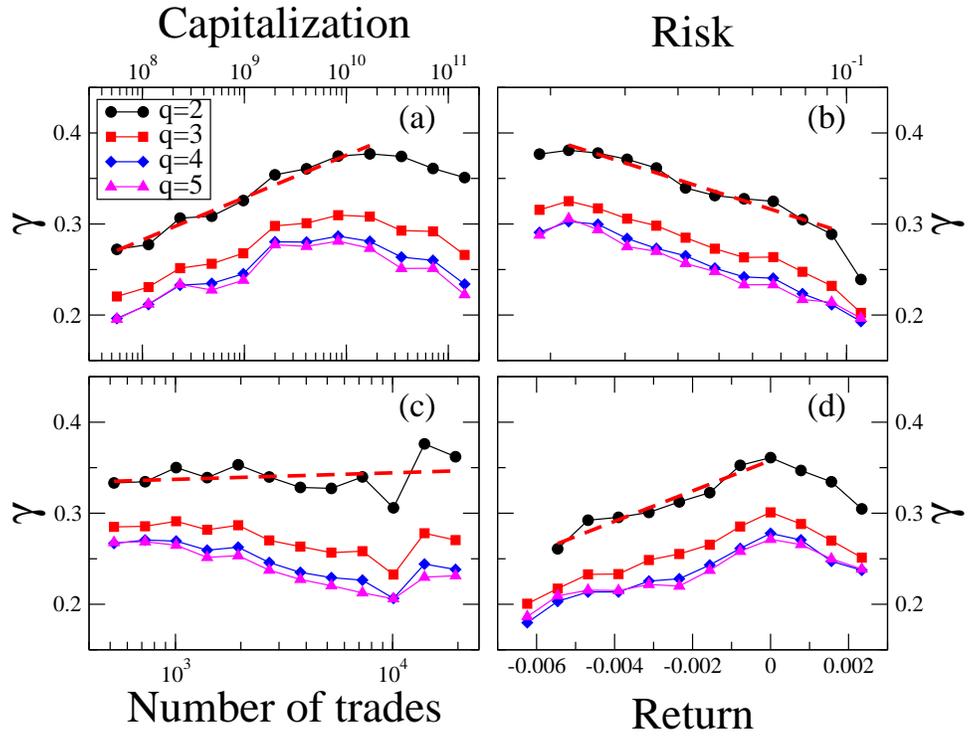}
\end{center}
\caption{(Color online) Relation between SE exponent $\gamma$ and four
factors: (a) market capitalization, (b) risk, the standard deviation
of daily return, (c) average daily number of trades and (d) average
daily return. Curves of four thresholds $q=2$, $3$, $4$ and $5$ are
demonstrated. Dashed lines are logarithmic fittings (except for (d)
where the fitting is linear) on the curve of $q=2$.}
\label{Fig3}
\end{figure*}

Next we study the relations between $\gamma$ and the four essential
factors, market capitalization, risk, number of trades and
return. This tests the universality of $\gamma$ over the entire
market. If $\gamma$ is sensitive to some factors, the market as one
system is not of uniscaling. Furthermore, the dependence (if exist)
may indicate some origins for the multiscaling found in return
interval distributions. In Fig. \ref{Fig3}, we plot $\gamma$ against
the four factors for four thresholds, $q=2$, $3$, $4$ and $5$. In each
panel, curves have similar tendency and the value of $\gamma$
decreases with $q$. Note that the curves are closer to each other for
large thresholds. This finding is consistent with Fig. \ref{Fig2},
which shows that the mean $\gamma$ decreases with $q$ and reaches
almost a constant value for large $q$. More important, Fig. \ref{Fig3}
exhibits that $\gamma$ for a given threshold is not uniformly
distributed with the factor values and thus the market is of
multiscaling nature.

For the company size (Fig. \ref{Fig3}(a)), $\gamma$ increases for sizes
between $5\times10^7$ to $2\times10^{10}$ dollars and then shows a
slight decrease. The market depth for small companies limits the size
of investors and those companies usually attract some specific types
of investors. Therefore corresponding strategies may be relatively
similar and the volatility series tends to be strongly correlated
having a small $\gamma$ \cite{Bunde04}. With increasing size, more
investors are involved, which may ``randomize'' the long-term
correlations in volatilities. When the company size reaches a certain
limit, the constitution of investor types may be relatively stable,
some common modes might dominate volatilities and therefore the
correlations become stronger and $\gamma$ decreases with the size.

Fig. \ref{Fig3}(b) shows that $\gamma$ decreases with the risk except
for very low risks. Fig. \ref{Fig2} shows that larger volatilities
tend to have smaller $\gamma$. Larger risk means that the probability
of larger volatilities is higher. Therefore, Fig. \ref{Fig3}(b) is
consistent with Fig. \ref{Fig2}. Price movement is realized by trades
and the temporal structure of the volatility probably relates to the
size of trades. Counterintuitively, $\gamma$ is almost not sensitive
to the market activity. Fig. \ref{Fig3}(c) suggests no apparent
dependence between $\gamma$ and the number of trades \cite{Note1}, see
also \cite{Ivanov04}. A possibly reason is that many investors do not
change their strategies only because of the dramatic change of trading
frequency. Next we show in Fig. \ref{Fig3}(d) the relation between
$\gamma$ and the return. For negative returns $\gamma$ increases and
decreases for positive returns. It has a maximum when the return is
$0$. This behavior suggests that the return is related to the size of
risk. For returns with large magnitude representing high volatilities,
the corresponding risk is relatively high and therefore $\gamma$ is
small (see Fig. \ref{Fig3}(b)).

\begin{figure*}
\begin{center}
   \includegraphics[width=0.65\textwidth, angle = -90]{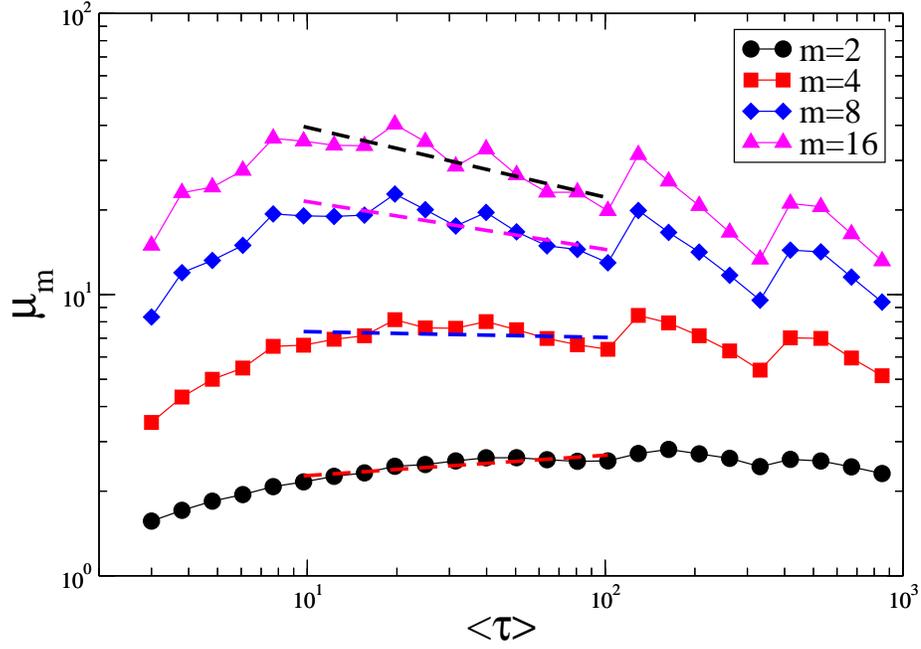}
\end{center}
\caption{(Color online) Typcial moments $\mu_m$ of the GE stock. Four
orders, $m=2$, $4$, $8$ and $16$ are shown. Dashed lines are power-law
fittings in the range of $10<\langle\tau\rangle\le100$ for determining
the multiscaling exponent $\delta$.}
\label{Fig4}
\end{figure*}

Next we study the multiscaling behavior of individual stocks. The
moments of scaled interval, $\tau/\langle\tau\rangle$, can quantify
the deviations from a single scaling function and therefore provide a
good measure to test the multiscaling in individual stocks. In
Fig. \ref{Fig4}, we plot $\mu_m$ for GE as an example. The moment and
the corresponding exponent $\delta$ for the multiscaling are defined
as
\begin{equation}
\mu_m \equiv \langle(\tau/\langle\tau\rangle)^m\rangle^{1/m} \sim
\langle\tau\rangle ^ \delta.
\label{delta.eq}
\end{equation}
If the distribution of the return intervals follows a unique scaling
law as Eq. (\ref{pdf.eq}), the different moments should be independent
on $\langle\tau\rangle$ and therefore the exponent $\delta$ should be
$0$. A significant $\delta$ suggests multiscaling, and the value of
$\delta$ characterizes the strength of the multiscaling \cite{Wang08},
thus we call $\delta$ {\it multiscaling exponent} \cite{Note2}. We
find that $\mu_m$ changes systematically with
$\langle\tau\rangle$. For $m \ge 2$, moments first increase with
$\langle\tau\rangle$ and then decrease \cite{Wang08} (for a typical
example, see Fig. \ref{Fig4}). Since a value of $\langle\tau\rangle$
corresponds to a threshold value $q$, the moments have the same trend
with $q$. Here we choose four typical orders for moments, $m=2$, $4$,
$8$ and $16$. For other positive orders, we find similar
behaviors. For these four orders, all 1137 stocks totally have 215 out
of 4548 cases ($4.7\%$) where the RMS error of fitting are over
$22\%$, and are not included in the analysis.

\begin{figure*}
\begin{center}
   \includegraphics[width=0.65\textwidth, angle = -90]{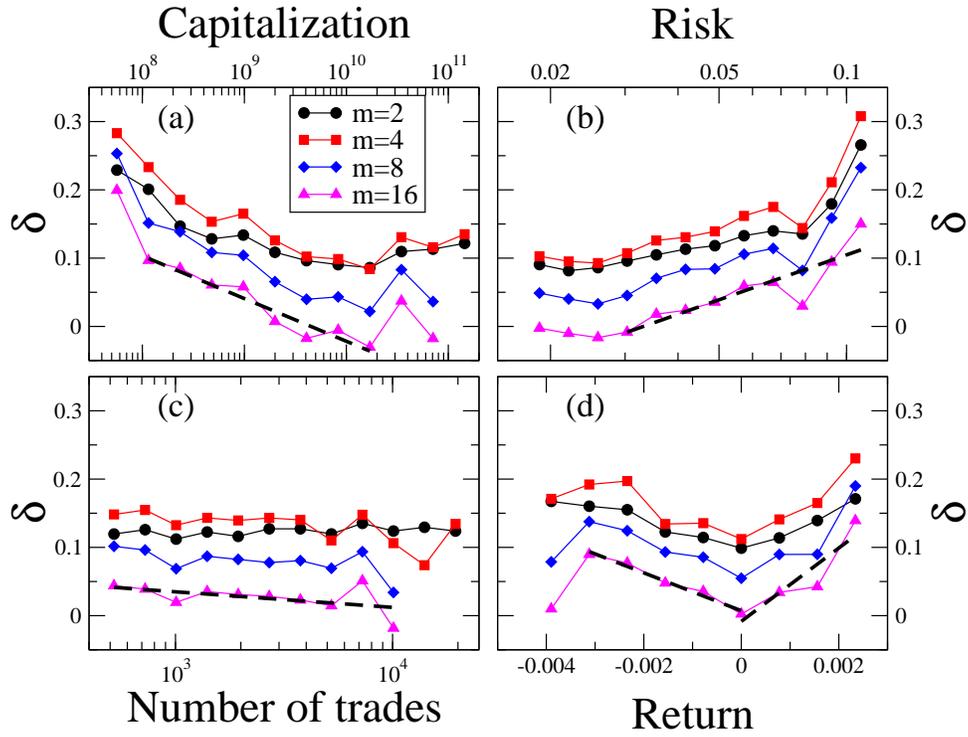}
\end{center}
\caption{(Color online) Relation between multiscaling exponent
$\delta$ and four factors: (a) market capitalization, (b) risk, the
standard deviation of daily return, (c) average daily number of trades
and (d) average daily return. Curves of four moments, $m=2$, $4$, $8$
and $16$ are shown. Dashed lines are fittings on the curve of $m=16$
which demonstrate the tendency.}
\label{Fig5}
\end{figure*}

Now we focus on the relation between the multiscaling exponent
$\delta$ and the four factors. We plot in Fig. \ref{Fig5} the curves
for four orders, $m=2$, $4$, $8$ and $16$. These curves have the
similar tendency in each panel. The value of $\delta$ increases a
little from $m=2$ to $4$, then decreases, which is consistent with the
result in Ref \cite{Wang08}. As shown in Fig. \ref{Fig5}(a), $\delta$
decreases with the capitalization until about $2 \times 10^{10}$
dollars and then the curves increase. This suggests that $\delta$ also
relates to the constitution of investors. A small company has few
investors which have some specific strategies. However, if the company
is very large, some types of investors finally dominate the price
movement. In Fig. \ref{Fig5}(b), $\delta$ increases almost
monotonically with the risk, indicating that if a stock has larger
volatility values, its return interval distribution has stronger
multiscaling effect. Similar to $\gamma$, $\delta$ is almost
independent on the number of trades as shown in Fig. \ref{Fig5}(c). In
Fig. \ref{Fig5}(d), $\delta$ has a minimum at zero returns, which also
agrees with the relation between $\delta$ and risk.

\begin{figure*}
\begin{center}
   \includegraphics[width=0.65\textwidth, angle = -90]{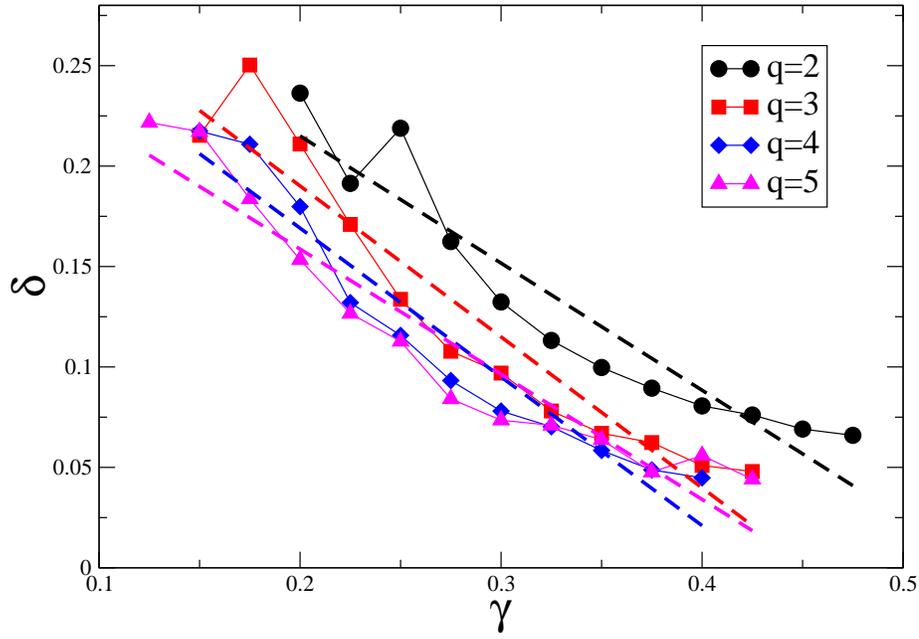}
\end{center}
\caption{(Color online) Multiscaling exponent $\delta$ vs. SE exponent
$\gamma$. The values of $\delta$ are for order $m=2$ and $\gamma$ are
for thresholds $q=2$, $3$, $4$ and $5$. Linear fittings are shown by
dashed lines.}
\label{Fig6}
\end{figure*}

There are clear connections between Fig. \ref{Fig3} and Fig
\ref{Fig5}, which indicate that $\gamma$ and $\delta$ are strongly
related. From Fig. \ref{Fig2}, $\gamma$ decreases with $q$, a $q$
value corresponds to a $\langle\tau\rangle$ value, and $\delta$ is the
power-law fitting exponent for the moment vs. $\langle\tau\rangle$. To
examine the relation between the two exponents we plot $\delta$ against
$\gamma$ in Fig. \ref{Fig6}. Our results suggest that $\delta$
decreases with $\gamma$ for all four thresholds $q=2$, $3$, $4$ and
$5$ when $m=2$. These curves approximately follow a linear function as
guided by the dashed lines with the slope $-0.63$, $-0.75$, $-0.74$
and $-0.62$ respectively. Other thresholds $q$ and orders $m$ show
similar results. The smaller is the value of $\gamma$, the larger
deviation from a single scaling function for the return interval
distribution is observed.

The SE exponent $\gamma$ characterizes the return intervals, which
depend on the temporal structure of volatility time series. In other
words, $\gamma$ characterizes the dynamic property of
volatility. Capitalization, risk and return are fund mental measures
of a company while number of trades is for the market activity, which
is due to the market participants and not influenced by the company
managers. We show that $\gamma$ relates to these fundamental measures
but not to the activity. We also test the relation between $\gamma$
and share volume, and find no clear dependence, similar to that for
number of trades. Although there is a certain relation between those
measures, for instance, the number of trades depends on the
capitalization \cite{Eisler06}, it does not guarantee that $\gamma$
depend on the number of trades. For a company of a given number of
trades, its capitalization has a range of values, and for a company of
a given capitalization, its $\gamma$ also distributes in a certain
interval. Since there is a crossover in the curve of $\gamma$ and
capitalization, it is possible that $\gamma$ is not sensitive to the
number of trades. Capitalization, risk and return are widely used for
building portfolio. Therefore, $\gamma$ connects the dynamic structure
of the price movement with fundamental measures, which may provides an
helpful indicator for portfolio selection. Similarly, the multiscaling
exponent $\delta$ also could be used to optimize the
portfolio. Recently Bogachev et al. studied return intervals in
multifractal data sets and suggested that the return interval follows
a power-law distribution \cite{Bogachev07}. Meanwhile, Livina et
al. suggested a Gamma distribution for earthquake time series which
also has long-term correlations\cite{Livina05}. Therefore a detailed
analysis on the distribution function is needed.

In summary, we analyzed the volatility return interval for 1137 most
traded stocks in the United States markets. We have shown that the SE
exponent $\gamma$ depends on the threshold $q$, which supports
multiscaling nature in the return interval distribution. We also
studied the relation between $\gamma$ and four essential factors of
stock, capitalization, risk, number of trades and return. We found
that $\gamma$ depends on the capitalization, risk and return but not
on the number of trades, which suggests the multiscaling in the entire
market. We further analyzed the multiscaling exponent $\delta$, which
characterizes the multiscaling of individual stocks. We found that it
again depends on the capitalization, risk and return but not on the
number of trades. Our results suggest that $\delta$ and $\gamma$ may
be useful for portfolio optimization.

We thank S.-J. Shieh, R. Mantegna, J. Kert\'{e}sz and Z. Eisler for
helpful discussions, and the NSF and Merck Foundation for financial
support.

\newpage

\end{document}